\documentstyle[aps,prb,multicol,psfig]{revtex} 
\newif\ifmynarrow
\mynarrowfalse
\renewcommand{\narrowtext}{%
  \ifmynarrow\hspace*{\fill}\raisebox{-1ex}[0pt][0pt]{%
    \rule{0.3pt}{1ex}%
    \rule[1ex]{20.5pc}{0.3pt}}\fi \mynarrowtrue
  \vspace{-1.0ex}%
  \begin{multicols}{2}%
  \par\global\columnwidth20.5pc
  \global\hsize\columnwidth\global\linewidth\columnwidth
  \global\displaywidth\columnwidth}
\renewcommand{\widetext}{%
  \end{multicols}%
  \vspace{-2.5ex}%
  \noindent\raisebox{1ex}[0pt][0pt]{%
    \rule{20.5pc}{0.3pt}%
    \rule{0.3pt}{1ex}}%
  \par\global\columnwidth42.5pc
  \global\hsize\columnwidth\global\linewidth\columnwidth
  \global\displaywidth\columnwidth}
\begin{document}
\draft 
\tighten

\title{Kondo correlation and spin-flip scattering in spin-dependent transport\\ 
through a quantum dot coupled to ferromagnetic leads} 
\author{Jing Ma and X. L. Lei}
\address{Department of Physics, Shanghai Jiaotong University,
1954 Huashan Road, Shanghai 200030, China}
\date{\today}
\maketitle

\begin{abstract}
We investigate the linear and nonlinear dc transport through an interacting
quantum dot connected to two ferromagnetic electrodes around Kondo regime with 
spin-flip scattering in the dot. Using a slave-boson mean field approach for
the Anderson Hamiltonian having finite on-site Coulomb repulsion, we find that 
a spin-flip scattering always depresses the Kondo correlation at arbitrary 
polarization strength  in both parallel and antiparallel alignment of the lead
magnetization and that it effectively reinforces the tunneling related
conductance in the antiparallel configuration. For systems deep in the Kondo 
regime, the zero-bias single Kondo peak in the differential conductance is split
into two peaks by the intradot spin-flip scattering; while for systems somewhat 
further from the Kondo center, the spin-flip process in the dot may turn 
the zero-bias anomaly into a three-peak structure.
 
\end{abstract}

\narrowtext

Spin-dependent electron transport through systems consisting of 
two ferromagnetic layers (FM) sandwiched by a quantum-dot (QD) has recently become
one of the major focuses of the rapidly developing spin-electronics.\cite{Prinz}    
Different from other nonmagnetic sandwiches, a quantum dot features 
its small size and the strong many-body correlation among electrons in it such that, 
not only the Coulomb blockade can dominate its electronic transport,
but particularly, a significant Kondo effect may arise. Very rich 
behavior has been disclosed when varying the strength and/or the relative orientation 
of the spin polarizations of two magnetic electrodes, the dot energy level 
and the on-site Coulomb correlation in the dot.\cite{WR,FM,NS,JM,JK,RL,MA,PZ,RLO} 
Effect of spin-flip scatterings in the QD has also been 
explored very recently. By means of the equation of motion method\cite{Meir} 
and the Ng ansatz\cite{Ng}, Zhang {\it et. al.}\cite{PZ} predicted that, 
in a FM/QD/FM system having infinite on-site Coulomb repulsion, 
the presence of the spin-flip process in the dot may split the original 
single Kondo resonance in the density of states into two or
three well-defined peaks. Utilizing a slave boson mean field (SBMF) technique 
for infinte-$U$ systems,\cite{Cole,Agu} L\'{o}pez and S\'{a}chez\cite{RLO} found 
a splitting of the zero-bias peak of the nonlinear differential conductance 
when the spin-flip scattering amplitude is of the order of the Kondo temperature.
However, the failure of detecting the Kondo peak splitting induced by  
lead magnetization in the parallel configuration and the unphysical prediction 
of complete quenching of the zero-bias anomaly and vanishing widths of 
the nonzero-bias peaks at large spin-flip scattering,\cite{RLO} indicate the 
limitation of the infinite-$U$ SBMF  approach to FM/QD/FM systems.  

In this Letter we investigate effects of intradot spin-flip scatterings  
on the dc conductance through a FM/QD/FM system using the finite-$U$ SBMF approach 
of Kotliar and Ruckenstein.\cite{KR,Dor,Dong1}  With the help of 
four slave boson parameters,
this SBMF approach not only allows to deal with finite on-site Coulomb
repulsion but also to take account of charge fluctuations to certain degree.
For a FM/QD/FM system without spin-flip scattering\cite{MA} it derives 
a clear splitting 
of the Kondo peak in differential conductance due to lead magnetization 
in the parallel configuration in agreement with other techniques,\cite{JK,RL} 
and predicts the physically reasonable peak widths and heights. 

We consider a quantum dot (QD) of single bare level $\epsilon_d$ having a
finite on-site Coulomb repulsion $U$ and a spin-flip scattering amplitude $R$, is
coupled to two ferromagnetic leads through tunneling $V_{k_{\alpha}\sigma}$
($\alpha=L, R$). Both the left and the right leads are magnetized 
along the $z$ axis but may be in parallel (P) or antiparallel (AP) alignments 
(configurations) 
and the electrons in the leads are described by wavevector $k$ and spin index
$\sigma$ with energy $\epsilon_{k_{\alpha}\sigma}\,(\sigma=\pm 1 \, {\rm or} \uparrow 
\downarrow)$. 

In terms of the finite-$U$ slave-boson approach,\cite{KR}
we introduce four auxiliary boson operators $e$, $p_\sigma$ 
and $d$, which are associated respectively with the empty, singly occupied and doubly 
occupied electron states of the QD, to describe the above physical problem without 
interparticle coupling in an enlarged space with constraints.
Within the mean-field scheme one can start with the following effective Hamiltonian:
\begin{eqnarray}
H_{eff}&=&\sum\limits_{k_\alpha\sigma}\epsilon_{k_\alpha\sigma}
c^{\dag}_{k_\alpha\sigma}c_{k_\alpha\sigma}
+\sum\limits_{\sigma}\epsilon_{d}c^{\dag}_{d\sigma}c_{d\sigma}+Ud^{\dag}d
\nonumber\\
&+&R(c^{\dag}_{d\uparrow}c_{d\downarrow}+{\rm H.c.})
+\sum\limits_{k_\alpha\sigma}[V_{k_{\alpha}\sigma}c^{\dag}_{k_\alpha\sigma}c_{d\sigma}
z_{\sigma}+{\rm H.c.}]\nonumber\\ 
&+&\lambda^{(1)}(\sum\limits_{\sigma}p^{\dag}_{\sigma}p_{\sigma}+e^{\dag}e+
d^{\dag}d-1)\nonumber\\
&+&\sum\limits_{\sigma}\lambda^{(2)}_{\sigma}(c^{\dag}_{d\sigma}c_{d\sigma}-p^{\dag}
_{\sigma}p_{\sigma}-d^{\dag}d),
\end{eqnarray}
where $c^{\dag}_{d\sigma}(c_{d\sigma})$ and
$c^{\dag}_{k_\alpha\sigma}(c_{k_\alpha\sigma})$ are the creation
(annihilation) operators of electrons in the dot and in the
electrodes, the three Lagrange multipliers $\lambda^{(1)}$ and
$\lambda^{(2)}_{\sigma}$ take account of the constraints, and in
the hopping term the fermion operator $c^{\dag}_{d\sigma}$ and $c_{d\sigma}$ are
replaced by $z^{\dag}_{\sigma}c^{\dag}_{d\sigma}$ and
$c_{d\sigma}z_\sigma$ to consider the many body effect on
tunneling. $z_\sigma$ embodys the physical process with which
an electron with spin $\sigma$ in the dot is annihilated:
$
z_\sigma=(1-d^{\dag}d-p^{\dag}_{\sigma}p_{\sigma})^{-\frac{1}{2}}(e^{\dag}p_\sigma+
p^{\dag}_{\bar{\sigma}}d)(1-e^{\dag}e-p^{\dag}_{\bar{\sigma}}p_{\bar{\sigma}})
^{-\frac{1}{2}}.
$
The Heisenberg equations of motion for the four slave-boson operators derived
from the effective Hamiltonian together with the three constraints, form the
basic equations. Then we use the mean-field approximation in the statistical 
expectations of these equations, in which all the boson operators are replaced 
by their expectation values which serve as the instrumental parameters 
in the mean-field scheme. In the wide-band limit the final equations are:
\begin{eqnarray}
\sum\limits_\sigma\frac{{\partial}z_\sigma}{{\partial}e}K_\sigma+2\lambda^{(1)}e=0,\\
\sum\limits_\sigma\frac{{\partial}z_\sigma}{{\partial}p_{\sigma'}}K_\sigma+
2(\lambda^{(1)}-\lambda^{(2)}_{\sigma'})p_{\sigma'}=0,\,\,\sigma'=\pm1,\\
\sum\limits_\sigma\frac{{\partial}z_\sigma}{{\partial}d}K_\sigma+2(U+\lambda^{(1)}-
\sum_\sigma\lambda^{(2)}_\sigma)d=0,\\
\sum\limits_{\sigma}|p_\sigma|^2+|e|^2+|d|^2=1,\\
\frac{1}{2{\pi}i}\int{d\omega}G^{<}_{d\sigma\sigma}(\omega)=|p_\sigma|^2+|d|^2,\,\,
 \sigma=\pm1.
\end{eqnarray}
Here
\begin{equation}
K_{\sigma}{\equiv}\frac{1}{\pi}{\int}d{\omega}\{z_{\sigma}{\rm
Re}(G^{r}_{d\sigma\sigma}(\omega))
[f_{L}(\omega)\Gamma_{L\sigma}+f_{R}(\omega)\Gamma_{R\sigma}]\},
\end{equation}
where $f_\alpha(\omega)=1/(e^{\beta(\omega-\mu_\alpha)}+1)$
is the Fermi function and $\mu_\alpha$ is the chemical potential 
of the $\alpha$th lead, which is assumed in an equilibrium state at temperature 
$1/\beta$, and
$\Gamma_{\alpha\sigma}(\omega)=
2\pi\sum_{k_\alpha}|V_{k_\alpha\sigma}|^2\delta(\omega-\epsilon_{k_\alpha\sigma})$
is the coupling strength function between the QD and the lead $\alpha$. 
$G^{r(a)}_{d\sigma\sigma'}(\omega)$ 
and $G^<_{d\sigma\sigma'}(\omega)$ are the elements of the $2{\times}2$
retarded (advanced) and correlation Green's function matrices
${\bf G}^{r(a)}_d (\omega)$ and ${\bf G}^{<}_d (\omega)$ in the spin
space of the QD. 
In the SBMF scheme, the strong correlating problem of the dot electrons with finite 
Coulomb repulsion $U$ is reduced to an effective noncorrelating one described 
by the effective Hamiltonian (1) with the slave-boson operators treated as c-numbers.
The retarded (advanced) Green's function can be expressed in the form renormalized
due to dot-lead couplings:
${\bf G}^{r(a)}_d (\omega)=\frac{1}{\omega{\bf I}-
\tilde{\bf H}_{d}\pm i\tilde{\bf \Gamma}}$,
in which 
$\tilde{H}_{d\sigma\sigma}=\tilde{\epsilon}_{d\sigma}$ 
($\tilde{\epsilon}_{d\sigma}=\epsilon_{d}+\lambda^{(2)}_\sigma$) 
and $\tilde{H}_{d\sigma\sigma^{\prime}}=R$ ($\sigma\neq\sigma^{\prime}$) 
take up the dot-level shifting and splitting and the spin-flip process,
$\tilde{\Gamma}_{\sigma\sigma'}=\frac{1}{2}|z_\sigma|^{2}(\Gamma_{L\sigma}+
\Gamma_{R\sigma})\delta_{\sigma\sigma'}$ reflects the dot-level broadening, 
and ${\bf I}$ is a unit matrix.
The correlation Green's function ${\bf G}^{<}_d$ can be obtained by the Keldysh 
equation: $
{\bf G}_d^{<}={\bf G}_d^r {\bf \Sigma}_{d}^{<}{\bf G}_d^a,
$
with ${\Sigma}^{<}_{d\sigma\sigma'}=i|z_\sigma|^2(f_{L}\Gamma_{L\sigma}+
f_{R}\Gamma_{R\sigma})\delta_{\sigma\sigma'}$.

The electric current flowing from the left lead into the QD is obtained from
the rate of change of the electron number operator of the left
lead:\cite{Meir,Jauho}
\begin{eqnarray}
I_L=\frac{ie}{\hbar}\int\frac{d\omega}{2\pi}\sum\limits_{\sigma}\Gamma_{L\sigma}
|z_\sigma|^2[(G^{r}_{d\sigma\sigma}-G^{a}_{d\sigma\sigma})f_{L}+G^{<}_{d\sigma\sigma}].
\end{eqnarray}
In the steady transport state, the current flowing from the QD to the right lead must 
be equal to the current from the left lead to the QD, and the formula (8) can be 
directly used for calculating the current flowing through the lead-dot-lead system 
under a bias voltage $V$ between the two leads:
$I=I_L$. 

 We assume that the left and right leads are made from identical materials and 
that, in the wide band limit, the effective coupling strength functions are constants
for each direction of magnetization, such that in the P configuration 
$\Gamma_{R\uparrow}(\omega)=\Gamma_{L\uparrow}(\omega)=\Gamma_{\uparrow}$ 
and $\Gamma_{R\downarrow}(\omega)=\Gamma_{L\downarrow}(\omega)=\Gamma_{\downarrow}$,
and in the AP configuration
$\Gamma_{R\uparrow}(\omega)=\Gamma_{L\downarrow}(\omega)=\Gamma_{\downarrow}$ and
$\Gamma_{R\downarrow}(\omega)=\Gamma_{L\uparrow}(\omega)=\Gamma_{\uparrow}$.  
We will take $\Gamma\equiv(\Gamma_\uparrow+\Gamma_\downarrow)/2$
as the energy units and define the spin polarization as
$P\equiv (\Gamma_\uparrow-\Gamma_\downarrow)/(2 \Gamma)$. The Kondo temperature for 
finite-$U$ system in the case of $P=0$, given by
 $T_K^0=U\sqrt{\beta} \exp(-\pi/\beta)/2\pi$ 
 with $\beta=-2U\Gamma/\epsilon_d(U+\epsilon_d)$,
will be used as a reference dynamical energy scale.
In the following we will study FM/QD/FM systems having a fixed finite Coulomb
repulsion $U=6$ and concentrate on the effects of 
the spin-flip scattering on linear and nonlinear transport.
In calculating the dc current under a bias voltage $V$ between 
two leads, we choose the chemical potential 
$\mu_L=-\mu_R=eV/2$ for the left and right leads.

The finite-$U$ SBMF method was known to yield qualitatively correct conduction behavior 
at low temperatures within the bare dot-level range $-1.2U\leq \epsilon_d \leq 0.2U$.    
Fig.\,1 shows the calculated zero-temperature linear conductance $G=(dI/dV)_{V=0}$ 
as a function of $\epsilon_d$ of the QD having 
spin-flip scattering of different amplitude $R$ at fixed polarization $P=0.7$ 
in AP and in P configurations.
 
In the case of $R=0$, the $G$-vs-$\epsilon_d$ curve contains three regimes, covering 
the resonance peak due to the dot level around $\epsilon_d=0$, the charging peak around 
$\epsilon_d=-U$, and the Kondo peak centered at $\epsilon_d=-U/2$. 
In the AP configuration, all the renormalized parameters are identical for up 
and down spin indices at arbitrary $P$ and electrons with up-spin and down-spin 
are equally available in the whole lead-dot-lead system, favoring the formation 
of the Kondo-correlated state within a relatively wide dot-level range centered at 
$\epsilon_d=-3$. 
The $G$-vs-$\epsilon_d$ curves appear to be gentle-top hump structure.  
On the other hand, in the AP configuration the available minority-spin 
(e.g. up-spin) states in the right lead decline due to $P>0$ and 
the transfer of the majority-spin (up-spin) electrons from the left lead to 
the right lead is suppressed by the finite polarization, such that the conductance 
of the system goes down with increasing polarization from $P=0$ and vanishes at $P=1$.
The effect of a finite polarization is to reduce the height of $G$-vs-$\epsilon_d$ 
hump, while the shape of the curve remains essentially unchanged. 
\begin{figure}[htb]
  \psfig{figure=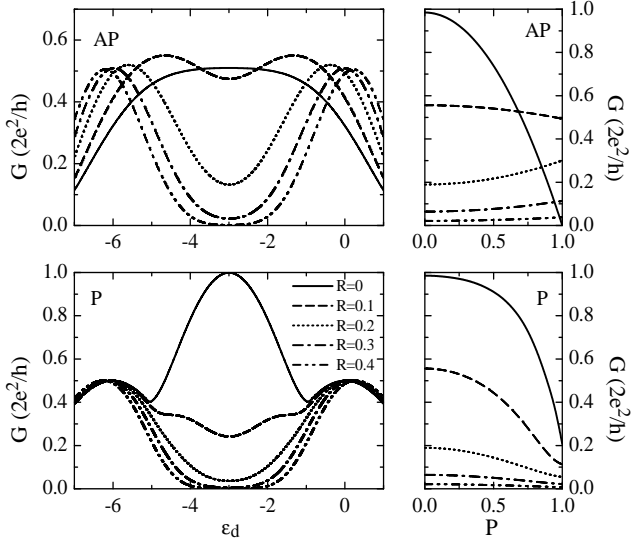,width=8.7cm,height=7.7cm,angle=0,clip=on}
    \caption{The linear conductance $G$ versus $\epsilon_d$ at $P=0.7$ (left) and 
versus $P$ at $\epsilon_d=-2$ (right) in AP and P configurations for different 
values of $R$. The systems have a finite value $U=6$.}
\end{figure}
In the P configuration, finite polarization reduces the available minority-spin
electrons in both electrodes and splits the dot level for up and down spins, thus 
weakens the Kondo correlation (narrowing the Kondo peak centered at $\epsilon_d=-3$) 
and slightly broadens the tunneling peaks around $\epsilon_d=0$ and $\epsilon_d=-U$. 
However, since both up- and down-spin electrons can tunnel through from the left lead 
to the right lead, the heights of the Kondo peak and tunneling peaks are essentially 
not affected by the finite polarization and the unitary limit
($G=2e^2/h$) can still be reached as long as $P<1$. 
The main feature in the P configuration is the central Kondo peak  
progressively narrowed with increasing $P$. 

A coherent spin-flip scattering ($R>0$) in the dot plays three important roles. 
(1) It lifts the degenerate dot level $\epsilon_d$ to $\epsilon_d \pm R$ 
even in the presence of a strong on-site Coulomb repulsion, with states being 
the linear combinations of the up- or down-spin ones. 
This level splitting may lead to somewhat broadening of the resonance peak 
and charging peak. (2) It strongly suppresses 
the Kondo correlation, such that the conductance maximum at $\epsilon_d=-3$, 
which is fully Kondo-induced, drops sharply when $R>0$. (3) In the AP configuration, 
$R>0$ makes it possible for the majority-spin electrons of the left lead to change 
their spin orientations after tunneling into the dot, in favor of their continuing 
 to tunnel into the right lead, and thus enhances the tunneling related conductance. 
 As can be seen in Fig.\,1,
the tunneling peaks and the charging peaks in the cases of $R=0.1,0.2,0.3$ 
and 0.4 are all higher than those of $R=0$. In the P configuration,
however, the tunneling-related conductance is essentially not affected 
by the presence of spin-flip process. The $G$-vs-$\epsilon_d$ curves in Fig.\,1 
are the result of competition and compensation of the above three effects.           
We see that, in the P configuration, the spin-flip scattering reduces the 
central Kondo peak sharply while slightly shifts the resonance peak and charging 
peak apart, finally turning the three-peak $G$-vs-$\epsilon_d$
curve into a Coulomb blockade type double-peak one. In the AP 
configuration, a deep valley develops progressively at the center $\epsilon_d=-3$
from the previous $G$-vs-$\epsilon_d$ hump, leading also to a Coulomb blockade 
type double peak one with enhanced heights of and enlarged distance between
two peaks.

Effects of changing polarization strength on linear conductance in the presence of 
spin-flip process are shown in the right part of Fig.\,1
for the system of $\epsilon_d=-2$ in the Kondo regime. In the P configuration,
$G$ always goes down with growing $P$ for fixed $R$ and with growing $R$ for fixed
$P$, indicating that both the polarization and the spin-flip scattering
suppress the Kondo resonance. In the AP configuration,
although the linear conductance $G$ still declines with increasing $P$ in the case of 
$R=0$ or with increasing $R$ in the case of $P=0$, 
anomaly shows up at finite $P$ and finite $R$ due to the competing effects of 
polarization and spin-flip scattering on Kondo correlation and resonance tunneling. 
At $R=0.1$ the conductance descends much slower than at $R=0$, and $G$ even 
rises with increasing polarization for $R\ge 0.2$. 
\begin{figure}[htb]
  \psfig{figure=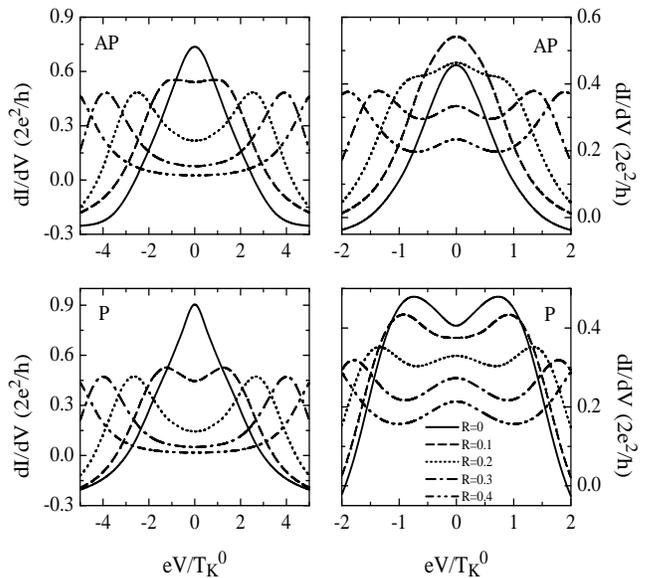,width=8.7cm,height=8.0cm,angle=0,clip=on}
  \caption{Zero-temperature $dI/dV$ versus $eV/T_K^0$ for $\epsilon_d=-2$ at $P=0.5$ 
(left) and for $\epsilon_d=-1$ at $P=0.7$ (right) having different values of spin-flip 
amplitude $R$ in AP and P configurations. The systems
have a finite value $U=6$.}
 \end{figure}
Zero-temperature differential conductance $dI/dV$ is calculated at bias voltage $eV$ 
up to several $T_K^0$. Fig.\,2 left shows  $dI/dV$ versus $eV/T_K^0$
at different spin-flip scattering amplitudes 
$R=0,0.1,0.2,0.3$ and $0.4$ but same polarization $P=0.5$ for the system 
of $\epsilon_d=-2$ in the Kondo regime ($T_K^0=0.144$). 
For this system at $P=0.5$
in the absence of spin-flip process, the dot-level splitting induced 
by the lead polarization alone in the P configuration has not yet been able  
to split the broadened Kondo peak. We have essentially the conventional
zero-bias anomaly of Kondo correlation in both AP and P configurations for $R=0$. 
Turning on the intradot spin-flip scattering suppresses the Kondo effect and 
lifts the dot-level degeneracy of order of $R$, giving rise to a significant drop 
of the height of the zero-bias anomaly and, up from $R=0.1$ ($\sim 0.7\,T_K^0$),
$dI/dV$ begins to split into two peaks centered around $eV=\pm 2R$ in both 
configurations.
This result is in agreement with the recent prediction of L\'{o}pez and
 S\'{a}nchez.\cite{RLO} Note that, however, our $dI/dV$ for this system does not 
show any sign of complete quenching of the zero-bias conductance and 
the split peaks predicted here exhibit physically reasonable widths 
of order of renormalized level broadening.

Situation becomes somewhat different for the system of $\epsilon_d=-1$,
a little further from the center of the Kondo regime,
as shown in the right part of Fig.\,2. For this system at $P=0.7$, in the P 
configuration the lead-polarization induced dot-level splitting is already large enough 
to show up as the zero-bias peak splitting
in the case of $R=0$. Introducing spin-flip scattering further reduces the zero-bias
differential conductance and enlarges the separation between the two split peaks 
in such a way that the distance increment is roughly 4 times of the R increment except 
the case from $R=0$ to $R=0.1$.
Note that for this system the zero-bias conductance, which is already much lower than
the unitary limit $2e^2/h$ due to the suppression of lead magnetization, comes only 
partly from Kondo correlation and partly from resonance tunneling. The spin-flip 
process suppresses the Kondo contribution while keeps the latter essentially
unchanged. Therefore the drop of the zero-bias $dI/dV$ is modest in comparison 
with that in the left part of Fig.\,2. On the other hand, the peak splitting develops 
progressively due to the increase of $R$-induced level splitting. 
The appearance of three-peak structure in  
the curves of $R=0.2,0.3$ and 0.4 is just the result of these two effects.
In the AP configuration, there is no dot-level splitting and the differential 
conductance remains a zero-bias single-peak curve for any strength
of the lead polarization in the case of $R=0$. The spin-flip scattering, while 
suppressing the Kondo contribution, helps to
enhance the resonance-tunneling related conductance. For $\epsilon_d=-1$
system the latter overcompensates the Kondo reduction, such that we see  
the whole $dI/dV$ curve of $R=0.1$ rising over that of $R=0$. Note that 
at $R=0.1$, the spin-flip induced level splitting has not yet been able 
to destroy the single-peak structure and we have only a broadened peak. 
The $R=0.2$ induced level splitting leads to the appearance of shoulders 
on both sides. Two side peaks clearly show up at larger spin-flip scattering 
amplitudes. 
We see three-peak $dI/dV$ curves for $R=0.3$ and 0.4 in the AP configuration.

In summary, we have demonstrated that a spin-flip scattering in QD 
always suppresses the Kondo correlation at arbitrary 
polarization strength in both P and AP alignment of the lead magnetizations, 
and the effect is stronger for P configuration at larger $P$. 
It effectively reinforces the tunneling related
conductance in the AP configuration. For systems deep in the Kondo 
regime, the zero-bias single Kondo peak in the differential conductance is split
into two peaks by the intradot spin-flip scattering; 
while for system somewhat further from the symmetric point, the spin-flip process 
in the dot may turn the single-peak conductance curve into a three-peak form. 
The predicted positions and widths of the split peaks are  
in rough agreement with the spin-flip induced energy splitting 
and the tunneling induced level broadening, renormalized by the many-body
effect. The main physics concerned in this Letter is in the energy scale of the Kondo
temperature $T_K^0$ or the effective coupling strength $\Gamma$, which are easily 
accessible experimentally.

We would like to thank Drs. Bing Dong and S.Y. Liu for helpful discussions.
This work was supported by the National Science Foundation of China,
the Special Funds for Major State Basic Research Project, and
the Shanghai Municipal Commission of Science and Technology.

\end{multicols}
\end{document}